# THE METHOD OF DETECTING ONLINE PASSWORD ATTACKS BASED ON HIGH-LEVEL PROTOCOL ANALYSIS AND CLUSTERING TECHNIQUES

Nguyen Hong Son[1] and Ha Thanh Dung[2]

[1]Faculty of Information Technology
Posts and Telecommunications Institute of Technology, Vietnam
[2]Faculty of Information Systems and Remote Sensing
Ho Chi Minh City University of Natural Resources and Environments, Vietnam

*ABSTRACT*

*Although there have been many solutions applied, the safety challenges related to the password security mechanism are not reduced. The reason for this is that while the means and tools to support password attacks are becoming more and more abundant, the number of transaction systems through the Internet is increasing, and new services systems appear. For example, IoT also uses password-based authentication. In this context, consolidating password-based authentication mechanisms is critical, but monitoring measures for timely detection of attacks also play an important role in this battle. The password attack detection solutions being used need to be supplemented and improved to meet the new situation. In this paper we propose a solution that automatically detects online password attacks in a way that is based solely on the network, using unsupervised learning techniques and protected application orientation. Our solution therefore minimizes dependence on the factors encountered by host-based or supervised learning solutions. The certainty of the solution comes from using the results of in-depth analysis of attack characteristics to build the detection capacity of the mechanism. The solution was implemented experimentally on the real system and gave positive results.*

## 1. INTRODUCTION

The password-based protection mechanism is still in common use today despite the availability of biometric-based protection solutions. Therefore, the classic password attack is still a concern for many people and the attack is still raging on the Internet every day. Password attacks are often divided into several forms including brute force attacks, dictionary attacks, and rainbow table attacks. The password attack is done offline or online and the purpose of the attack is to obtain user's password and administrator password, thereby mastering the system. Thus password attacks are not only dangerous for personal accounts but also dangerous for web applications and databases. Protection against password attacks has become a topic of particular concern and a topic of new challenges. So far, many solutions have been applied from simple to complex, such as limiting the number of false attempts[4], using CAPTCHA, IP-based authentication blockers.

However, there have been many successful password attacks, as evidenced by the fact that many social network users who have lost their accounts or websites that have been hijacked. In the current password protection mechanism, every time a user enters a password into an Internet application's form, the password will be converted into a hash code and the hash code is sent to the server for comparison. Thus a lot of password hash codes are transmitted on the Internet from the user community. These hash codes are based on one-way functions and are confident that passwords can not be retrieved from them. Some hash algorithms are commonly used for password hashing such as MD5, SHA-1, SHA-2, SHA-256, and SHA-512. However, when the hashed password database is collected and an attempt to test with the assumed password from a given dictionary in turn, finding a hash code is real, we said the password was cracked. Unfortunately, such dictionaries are not so hard to find now. In addition, hackers do not have to manipulate manually to perform this test, but there are many software programs that serve as an effective support tool for attack activities, such as Cain and Abel, John the Ripper, RainbowCrack, which includes handy online attack tools like THC Hydra[17], Hashcat[18], Medusa, Ncrack, Brutus, and Patator.

Long and complex passwords with advanced hash functions will make password attacks more difficult and take more time.The hindrance of hackers will be mitigated if they have a high-powered computer to execute an attack program. The lifeline for hackers is that GPU virtual machines with large capacities are plentiful and easy to hire at a reasonable cost from many cloud providers. Besides, crypto currency technology is also designed to control a large amount of repetitive activity in mining that essentially detects the hash code. It can be said that crypto currency includes brute force technology. Both things just mentioned thus increase the threat from password attacks.

While many systems use password mechanisms, it is not convenient to implement attack protection mechanisms such as CAPTCHA, IP-based blockers. For example, FTP servers, FTP servers in IoT systems, content management systems (CMS) such as Wordpress, Drupal, Joomla, web hosting management systems (web panels) such as Plesk, Cpanel, and DirectAdmin.The systems all need to have effective protection against password attacks in the context above. Such systems often rely on host-based monitoring systems to detect online attacks but encounter problems with the consumption of host capacity, or difficulties in accessing to logs of protected systems. Thus, it makes host-based solutions more inadequate and unable to protect systems effectively. Since then we have found that the network-based online attack monitoring system is a viable solution to this problem.

In this paper, we will propose an approach to building an online password attack monitoring system that relies entirely on the network without the need for protected systems, which is not based on the host. First of all, we will propose a general model for construction with details of the essential steps to successfully build an attack monitoring application running on a standalone machine. In the model, it will exploit the strength of machine learning technology in unsupervised direction, using advanced data clustering techniques to separate different groups of subjects. The special thing of our proposal is to build an application-oriented attack detection application, which is to protect specific systems. On that basis we will combine with protocol analysis in traffic to pre-process data and build association rules to detect attacks. Another special feature is that we do not use clustering in which abnormal data does not belong to any cluster or to small clusters to detect abnormalities like other solutions often do [7] [9-10], in which [7] put abnormalities into small clusters. We aim to create clusters for attack data for analysis and detection. As an example of applying the solution, we have applied the model to build a password attack detection application that protects FTP servers and employs this application on an actual network to evaluate.

The rest of the paper is presented in the following sections: Section 2 is an overview of published studies that address similar problems and especially those that have solutions close to us. Some works related to the use of clustering techniques in recent times have also been generalized. Next in Section 3, we will propose a model to build an attack detection application along with detailed explanations of how to build components. Using the model to build the password attack detection application for the FTP server is shown in Section 4. Section 5 presents the deployment and experimentation of the attack detection application on actual network and results. The paper ends with some conclusions in Section 6.

## 2. RELATED WORK

Strengthening the password mechanism has always been a special topic of concern, which is reflected in many works proposing solutions to consolidate passwords [3][5-6][11][13-15]. In [13] the authors proposed a random password generation scheme using Markov chain techniques that are now known as Markov passwords. Markov passwords are generated according to a tree structure that creates the most powerful password. In [3] the authors also conducted experiments and evaluated the attack capability on Markov passwords and showed the effectiveness of this type of password. In [5] the authors propose a new cryptographic technique based on min-entropy key called Honey Encryption, which is believed to be able to resist brute force attacks.

However, practically no strong password mechanism can guarantee absolute safety against brute force attacks. Many studies have proposed solutions to overcome this classic problem. The authors in [16] consider using learning machines to detect brute force attacks via SSH protocol at the network level. The solution creates a monitoring tool that detects attack signals through traffic collected directly from the network layer. It makes the solution more scalable and so is called a network-based solution. However, their solution uses a supervised learning mechanism through the use of classifiers, so a good training data set is required. Similarly, the authors in [1] found that the limitations of host-based attack detection solutions are being implemented in most web application protection systems, thus suggesting a solution for detecting brute force attacks based on network. In the solution, the authors get traffics directly from the network environment and histogram the packet payloads on each network connection. Their proposal also uses HCA (Hierarchical Cluster Analysis) clustering technique to cluster based on the distance between clusters in the hierarchy. The detection phase is performed on large clusters, which will compare the histograms of connections if multiple identical connections appear, brute force attacks are present. In the solution, the histogram of connections is the main base of the detection but the construction of histograms and comparisons may lead to many errors. In addition, there are some network applications that repeat the connection but not the attack also give a significant amount of similar histogram.

Another attack detection solution also uses machine learning and does not rely on hosts, the authors in [7] also exploit unsupervised learning techniques to detect network attacks. The solution uses clustering techniques and incorporates correlation analysis to detect network attacks. The authors group anomalies into small clusters and implement multiple aggregation methods to analyse the correlation and rank the anomalies, thereby providing a security monitoring tool for administrators. Their proposal aims to detect all attacks and processing systems through multiple stages that need to be configured appropriately and verified.

Similarly, [9] also uses multivariate correlation and data mining clustering techniques to build a solution for detecting network-based DDoS attacks. The author in [19] uses an unsupervised machine learning algorithm to detect DDOS attacks in the incoming Internet data. However, the solution that the author proposes is a host-based DDoS attack detection model. Another aspect, in order to improve the efficiency of abnormal cluster detection, [12] point out that the lack of monitoring of historical factors of clustering results will lead to errors, and proposed the algorithm History-ORUNADA, an improvement of Real-time Online Unsupervised Network Anomaly detection Algorithm (ORUNADA), to monitor and evaluate clustering results over time. Concerning security for IoT systems at risk of brute force attacks, [2] highlighted the risk of brute force attacks on FTP server systems that are commonly used in today's IoT systems; especially pay attention to attacks from within. The authors in [2] have proposed solutions to effectively identify attack patterns based on time-sensitive statistical relationship (also the correlation method).

## 3. THE PROPOSED MODEL FOR DETECTING ONLINE PASSWORD ATTACKS

The requirements when designing a network-based brute force activity detection model is:

- Detection based on network traffic rather than based on host
- Specify the sources that are attacking brute force
- Discover online and early
- No original data source to learn
- Get to know the normal login activities

From the above requirements, the model will have to regularly monitor network traffic in the protected area, for example in an enterprise network environment. Since there is no data set, machine knowledge is not built from supervised learning, instead machine knowledge is built on the principle of unsupervised learning. To fulfill other requirements such as indicating the source of the attack, recognizing common logon operations or early attack detection, it is necessary to rely on special characteristics from the network traffic data and specify useful features for the application. Clustering is one of the important tasks in unsupervised learning and is commonly used in current anomaly detection problems. Based on the requirement analysis, the proposed model will use clustering techniques on network traffic data to serve the task of detecting brute force activities on the protected network area.

The proposed model is described in Figure 1. According to the model, there will be four general stages: collecting network traffic data on a protected environment according to the given time and cycle, data pre-processing and extracting characteristics, clustering on data only includes extracted features and cluster analysis for detection. It is just a look at the model as a normal process in unsupervised learning applications, but the model proposed here will have specific characteristics in stages to promote work efficiency.

First of all, the proposed model only captures selected packets according to the intended purpose in the network traffic data collection stage. The reasons are that it is necessary to reduce online processing data, limit the impact on resources and system performance when deployed, and the applications that need to be protected are fully known. So only collect traffic packets related to applications that need to be protected (with login session). In the data pre-processing and feature extraction process, the developer continues to analyse to select the features wisely. Thorough

International Journal of Computer Networks & Communications (IJCNC) Vol.11, No.6, November 2019

analysis of the evolution of network traffic in the interaction between the target system and the cyber client community, especially the login session, can infer which traffic characteristics are the best for the goals of the application. This means selecting features so that it is easy to separate the phenomena or objects that need to be discovered. This will be illustrated through a specific case we developed in the Section 4. In the next third stage, the data with $k$ selected features is clustered according to an appropriate clustering algorithm. Most solutions using clustering techniques to detect abnormalities will separate normal data into clusters and single abnormal data does not belong to any cluster. However, with the specificity of detecting online password attack traffic, this type of data point will appear much so it is not individual or alone and can form a cluster. Therefore, in order to know whether to attack or not, it is necessary to analyze clusters and in the proposed model each cluster will be taken to the final processing stage. At the final stage of the model, each cluster is analyzed and calculated the indicator parameters specified by the design. The indicator parameters are determined based on the behavioral analysis of the attack application in interaction with the target system. Those are specific parameters to recognize a certain attack activity. The final processing step is like a detection procedure based on supervised learning techniques. Because the detection program has been taught how the parameter specifies an attack. A standard test (such as a combination rule) will be conducted based on the indicator parameters to have a conclusion related to a brute force attack.

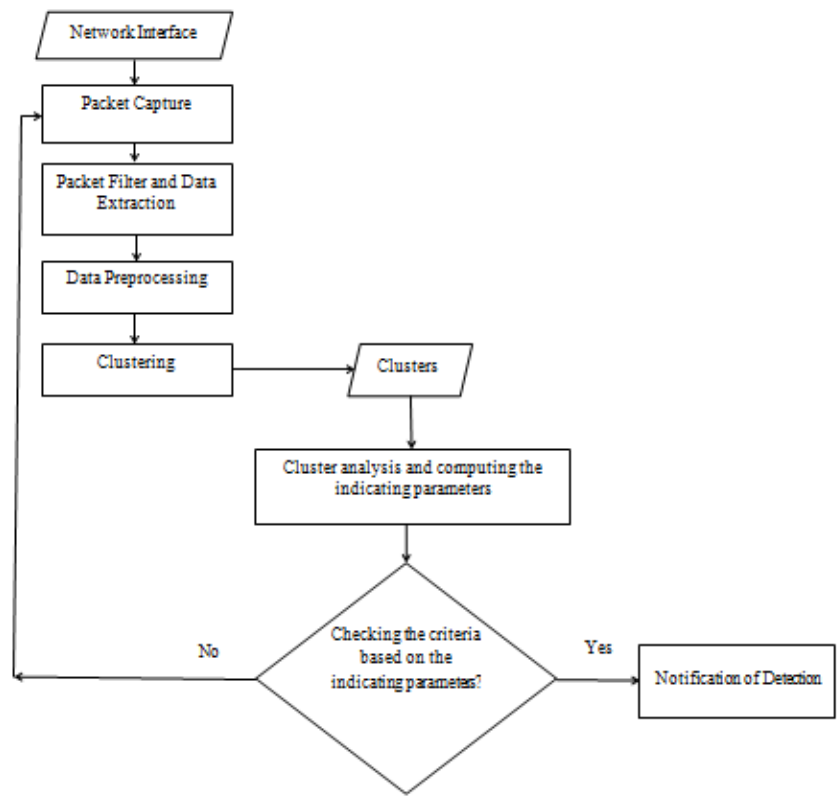

Figure 1. The proposed detection model

In the proposed model to build applications that detect brute force attacks, suppose the developer already knows protected target systems. This is entirely possible and very useful for developing fast and accurate attack detection applications. The basis of the proposal is that most network applications use one of the application layer protocols. It is possible for the packet capturing unit to retain the packets related to the application layer protocol used in the system that will be protected, and ignore unnecessary packets, contributing to reducing the amount of data processed in next step. In the phase of feature selection, it is also necessary to result from analysing the behaviour of the protocol since it will determine what is the unique characteristic of the traffic data generated by the corresponding application.

Finally, based on a thorough analysis of the behaviour of the protocol and the characteristics of the related attack, it is possible to identify certain indicators of the existence of attack operations that are hidden in the data. These indicators serve as signs that provide knowledge to the detection system used in the cluster analysis phase and to generate notifications. The address of the attacker will be determined by using a reverse analysis technique based on known attack data.

## 4. APPLYING THE PROPOSED MODEL TO DEVELOP AN APPLICATION THAT DETECT BRUTE FORCE ATTACK ON FTP SERVER

The function of the application is to monitor the campus network environment to detect and indicate the source of brute force attacks on the FTP server. The application has been developed based on the model proposed in Section 3. Accordingly, the first step is to capture the packet to retrieve the data and only capture packets related to the FTP protocol because the protection target is defined as the FTP server. Next will be selected characteristics and pre-processing of data collected before conducting clustering. The work at this stage is to analyze in-depth the traffic generated from the access and use of services on the FTP server. The initial FTP server connections are all directed to port 21. Both the normal FTP client and the offensive attack application undergo an attempt to log in via FTP protocol to port 21. Thus, the challenge here is no knowing where is the usual FTP client and where is the password attack. It is necessary to analyze the login session and extract the feature that has the specific nature of the brute force attack application to detect it. Based on packet capture and observation in Figure 2, it is found that each login session to the FTP server has 4 packets in the following order:

+ Request: USER
+ Response: 331 (Password required)
+ Request: PASS
+ Response: Status (530: Login failed, 230: Logged in successfully)

Through surveying the flow of brute force and non-brute force applications, the analysis on the data obtained has some of the following characteristics:

- Application layer protocol: This is the protocol that the application uses to transmit information in the application layer; this parameter cannot participate in calculations of the clustering algorithm.

- Transport layer protocol: Default is TCP, not involved in the calculation of the clustering algorithm.

| Time | Source | Destination | Protocol | Length | Info |
|---|---|---|---|---|---|
| 3 0.170870670 | 192.168.135.131 | 192.168.135.130 | FTP | 76 | Response: 530 Login incorrect. |
| 4 0.171140771 | 192.168.135.131 | 192.168.135.130 | FTP | 76 | Response: 530 Login incorrect. |
| 5 0.173417259 | 192.168.135.130 | 192.168.135.131 | FTP | 66 | Request: USER cuong |
| 7 0.173517295 | 192.168.135.131 | 192.168.135.130 | FTP | 88 | Response: 331 Please specify the password. |
| 8 0.173744577 | 192.168.135.130 | 192.168.135.131 | FTP | 66 | Request: USER cuong |
| 10 0.173796561 | 192.168.135.131 | 192.168.135.130 | FTP | 88 | Response: 331 Please specify the password. |
| 11 0.176144196 | 192.168.135.130 | 192.168.135.131 | FTP | 67 | Request: PASS 123635 |
| 12 0.176228288 | 192.168.135.130 | 192.168.135.131 | FTP | 67 | Request: PASS 123456 |
| 13 0.181623083 | 192.168.135.131 | 192.168.135.130 | FTP | 76 | Response: 530 Login incorrect. |
| 14 0.182073762 | 192.168.135.131 | 192.168.135.130 | FTP | 76 | Response: 530 Login incorrect. |
| 15 0.184830508 | 192.168.135.130 | 192.168.135.131 | FTP | 66 | Request: USER cuong |
| 17 0.185034719 | 192.168.135.130 | 192.168.135.131 | FTP | 66 | Request: USER cuong |
| 19 0.185565925 | 192.168.135.131 | 192.168.135.130 | FTP | 88 | Response: 331 Please specify the password. |

Figure 2. Traffic packets generated during the FTP server login session

- Port number: This is a port used for application in the transport layer, not involved in the calculation of clustering algorithm.

- Total number of packets from client-server: It is the number of packets in the direction from the client to the server of the connection (of an application) on the same access value during the collection time.

- Total TCP packets enable SYN flag: It is the number of TCP packets turning on the SYN flag, do not turn on the ACK flag of the connections (of an application) on the same access during the collection time.

- Time between two login sessions: It is the interval between two attempts to log in to a server during the recording time.

- The header length (hlen) of the packet created by the application.

- Time of each login session (unit of seconds): It is the time of communication between the client and the server in a login attempt.

- Number of bytes of each login session: It is the total amount of bytes generated during a login session.

```
Point ID    bytes    time(s)    hlen
point_0     305      3          5
point_1     306      3          5
point_2     329      3          5
point_3     353      0          8
point_4     353      1          8
point_5     324      0          8
point_6     353      0          8
point_7     353      0          8
point_8     354      0          8
point_9     354      0          8
point_10    354      0          8
point_11    354      1          8
point_12    354      0          8
point_13    354      0          8
point_14    354      0          8
point_15    354      0          8
point_16    354      0          8
point_17    354      0          8
point_18    352      1          8
point_19    352      0          8
point_20    352      0          8
...........
```

Figure 3. Clustering input data

Thus, in the above 9 features, the first 5 features can support the identification of normal applications by the specific protocol specified for the application or the port number is allocated separately for the application. However, for brute force attack applications, it is impossible to rely on these features to determine because there is no standard for this type of application. The solution is to combine the use of the last 4 features to cluster and attack prediction. Which will extract the last three characteristics for the clustering algorithm's input data and use the other feature to deduce the cluster is likely to be the application of brute force attack. Thus clustered data has three dimensions and takes the form shown in Figure 3.

In the third stage will cluster on data according to the appropriate clustering algorithm. In the stage, we propose to use the DBSCAN clustering algorithm. Before explaining the reason for choosing DBSCAN clustering algorithm, it is necessary to overview the characteristics of DBSCAN algorithm [8]. DBSCAN is a popular density-based clustering algorithm recently used to detect clusters and extract events. The basic idea of DBSCAN is that a point in the cluster must have number of neighbouring points at a given distance (Eps) not less than a MinPts value, which is the density. Based on such clustering, DBSCAN can divide clusters of any shape and users do not need to predetermine the number of clusters to be the algorithm's input parameter such as the K-mean algorithm. The advantage of DBSCAN is the ability to identify and distinguish very well among target groups. DBSCAN is particularly effective in spaces with few dimensions. Based on the analysis to develop the application to detect brute force attack on the FTP server mentioned above, use only 3 features for clustering so the clustering space has 3 dimensions. So applying the DBSCAN algorithm to clustering in this application is perfectly suitable.

The clusters created from DBSCAN will be analyzed and calculated the indicator parameters. In this application, based on the characteristics of the attack activity and normal operation, we choose the indicator parameter: the number of equal time intervals in the cluster. When an attack occurs, multiple logon sessions will fail and create points in the cluster with equal time series. This is the sixth feature of the selected features to extract a session connected to the FTP server shown above. When the indicator parameter reaches the threshold value which is a sign of an on-going attack and based on the information from the attack session cluster, the reverse analysis on the session data will determine the source of the attack.

## 5. TESTED ON ACTUAL NETWORK SYSTEMS AND RESULTS

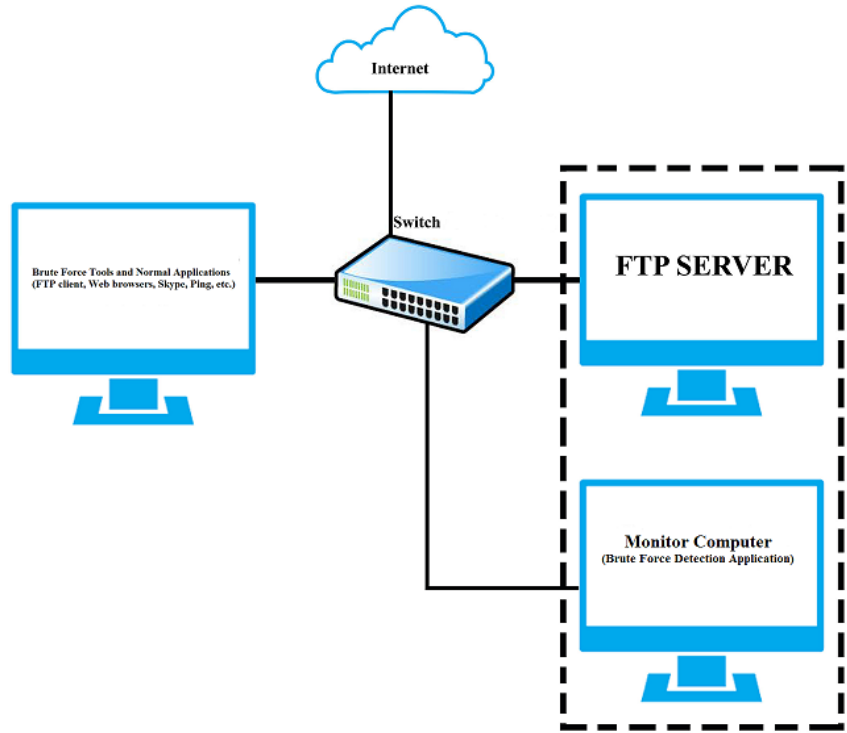

Figure 4. Network diagram for testing.

The application was developed using the Perl programming language and tested on our enterprise network, in principle, the network system is equivalent to the network described in Figure 4. The attack detection application running on a monitoring computer is located on the enterprise network partition so that it can catch the incoming and outgoing FTP server traffic. We experimented with continuous capture and processing of 1000 packets, with the clustering parameters Eps = 20 and MinPts = 10. Experiments were conducted in turn with the current popular attack tools including Hashcat, Brutus, Hydra, Patator, Medusa, and Ncrack. Other common applications are also run simultaneously to create a variety of traffic on the network.
The testing process will conduct each case as follows:

-Check the application to see if there is a mistake when running a normal application:

To test this case we use an FTP client to log in to the FTP server with a few failures before it succeeds and use ping to check the situation where the application repeats the session with the FTP server. As a result, both cases have no warning of attacks. As shown in Figure 5.

International Journal of Computer Networks & Communications (IJCNC) Vol.11, No.6, November 2019

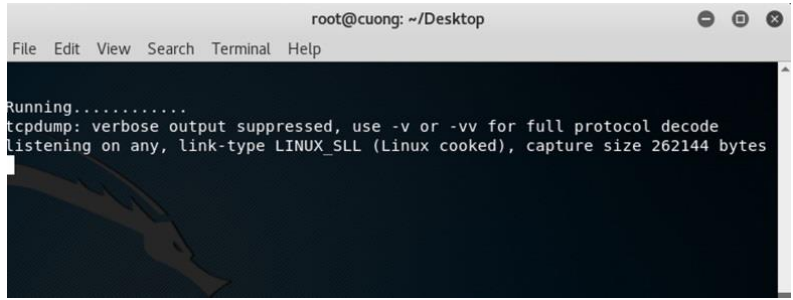

Figure 5. No attack warning.

- Check the ability to detect when attacking with different tools:

First attack the FTP server with the Medusa tool in the context of opening many other common applications like ping, Skype, web browser (YouTube), messenger. As a result of Figure 6, the application detected an attack and indicated the attack machine came from IP address 192.168.135.129 as it was.

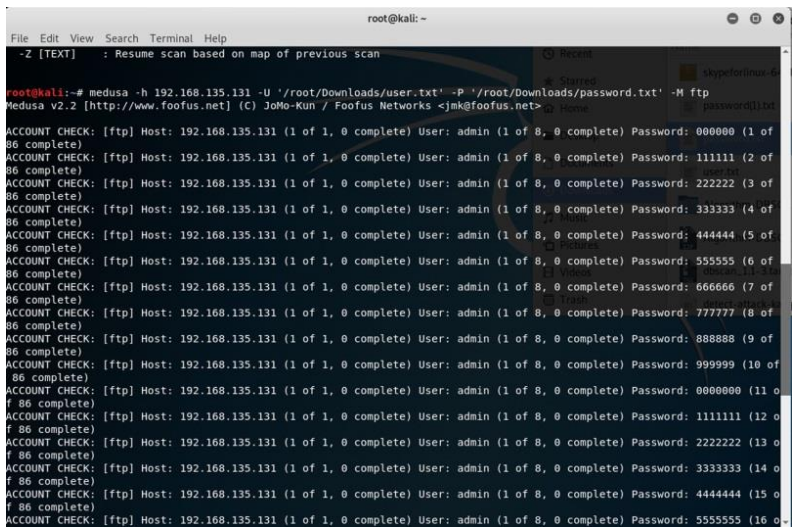

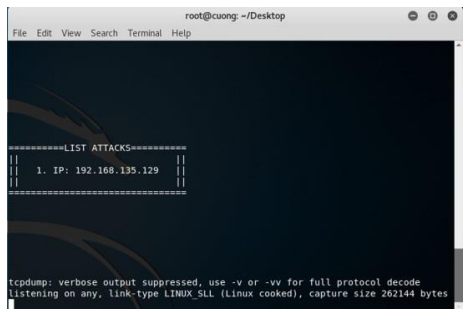

Figure 6. Detecting the attack machine address when attacking with the Medusa tool.

Next, turn to use Hashcat, Brutus, Patator, and Ncrack from 192.168.135.29 to attack the FTP server and run simultaneously with other applications. The results still detect and indicate the attack machine.

When using THC Hydra runs from two different machines to attack the FTP server, and also pings to the FTP server and runs various types of applications on other computers. The monitoring machine results show the IP addresses of the attacking machines as shown in Figure 7.

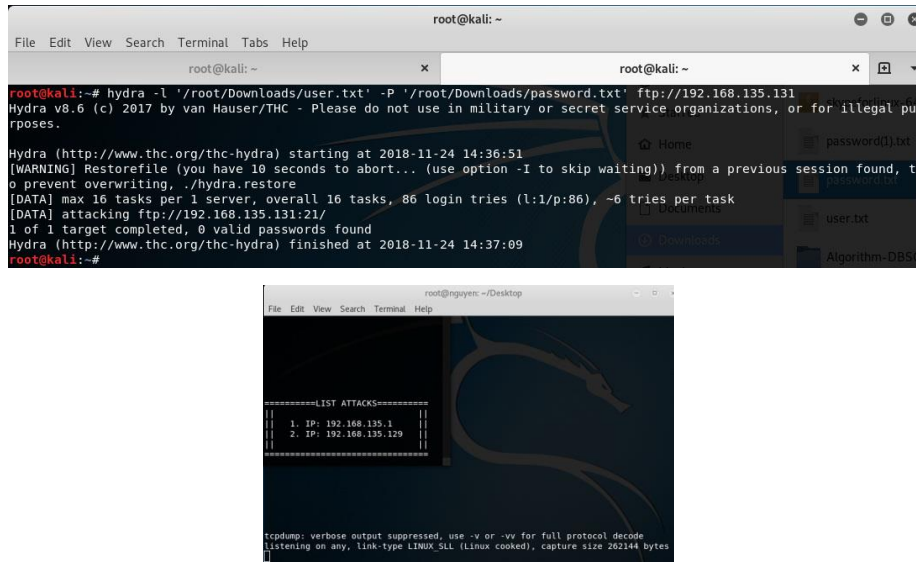

Figure 7. Detecting attack machine addresses when attacking with Hydra tool.

To evaluate the accuracy of the application, we conducted a test data set. This data set is based on collecting packet traffic from short-term automated attacks that are repeated many times randomly by four computers on the network over a 10 hours period. Tested on the data set with the change of clustering parameters, we tried many different Esp and MinPts parameter pairs. Based on understanding the data in the application we selected the Esp values in the range [10-80] and MinPts in the range [5-40] for testing. Some typical results are presented in Table 1.

The results show that the correct detection rate changes according to clustering parameters. High precision ratio corresponds to the case of Esp = 20 and MinPts = 10. When MinPts increases to 15 units, the accuracy decreases slightly. Especially when Esp doubled, the correct detection rate decreased significantly, and when both parameters increased, the rate decreased greatly.Thus, when choosing appropriate clustering parameters, the detected results can achieve almost absolute accuracy.

Table 1. The correct detection rates according to the clustering parameters

| **Esp** | **MinPts** | **True Positive** |
|---|---|---|
| 20 | 10 | 99% |
| 20 | 15 | 98% |
| 40 | 10 | 90% |
| 40 | 15 | 82% |
| 50 | 30 | 54% |

## 6. CONCLUSIONS

The password attack detection model with many new features has been proposed. The model aims to develop network-based automated detection applications that run on standalone machines that do not need to access protected host information. The model also aims to exploit advanced clustering techniques that can control any form of new password attack that occurs and does not require a training data set. In particular, the solution is oriented to the protected application, thus incorporating expert knowledge to protocol analysis to simplify the data before pretreatment, select the optimal characteristics in pre-processing, and build standards (rules) for attack detection effectively. Applying the model to build an application that detects password attacks on the FTP server has been implemented, and testing the application on the real system has given certain results. In the next research direction, we will improve the evaluation phase in which the experiment of protecting many different applications with high noise level. The results will be presented in subsequent publications.

International Journal of Computer Networks & Communications (IJCNC) Vol.11, No.6, November 2019[12] Juliette Dromard, Philippe Owezarski, (2017)"Integrating short history for improving clustering based network traffic anomaly detection", International Workshop on Autonomic Systems for Big Data Analytics (ASBDA 2017), Sep 2017, Tucson, United States. 8p. ffhal-01576752
[13] Vaithyasubramanaian.S and A. Christy, (2014) "A Scheme to Create Secured Random Password using Markov Chain", SPRINGER International Conference on Artificial Intelligence and Evolutionary Algorithm in Engineering Systems (ICAEES) 2014
[14] Jason Hong, Passwords Getting Painful, (2013) "Computing Still Blissful", Communications of the ACM March 2013, Vol.56, No. 3.
[15] Sarah Granger, (2002) "The Simplest Security: A Guideto Better Password Practices" http://www.symantec.com/connect/articles
[16] Maryam M. Najafabadi, Taghi M. Khoshgoftaar, Clifford Kemp, Naeem Seliya, and Richard Zuech, (2014) "Machine Learning for Detecting Brute Force Attacks at the Network Level", 2014 IEEE 14th International Conference on Bioinformatics and Bioengineering
[17] (2019) THC-HYDRA. [Online]. http://www.thc.org/thc-hydra/
[18] (2019) Hashcat. [Online]. http://hashcat.net
[19] Eric Perraud, (2019) "Machine Learning Algorithm of Detection of DoS Attacks on An Automotive Telematic Unit", International Journal of Computer Networks & Communications (IJCNC) Vol.11, No.1, January 2019
**AUTHORS**

**Nguyen Hong Son,** received his B.Sc. in Computer Engineering from The University of Technology in HCM city, his M.Sc. and PhD in Communication Engineering from the Post and Telecommunication Institute of Technology Hanoi. His current research interests include communication engineering, machine learning, data science, network security, and cloud computing.

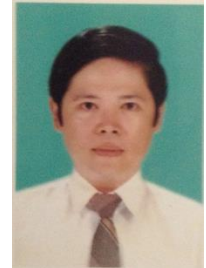

**Ha Thanh Dung**, received his B.Sc inInformation Technology from VNU Hanoi-University of Science, and his M.Sc in Data Transmission and Computer Networks from Post and Telecommunication Institute of Technology in 2012. His research areas are communication engineering, information systems, machine learning, and network security.

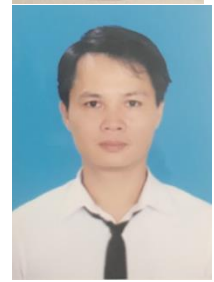
89